# Giant Phonon-induced Conductance in Scanning Tunneling Spectroscopy of Gate-tunable Graphene

Yuanbo Zhang[1*], Victor W. Brar[1,2], Feng Wang[1], Caglar Girit[1,2], Yossi Yayon[1], Melissa Panlasigui[1], Alex Zettl[1,2], Michael F. Crommie[1,2*]

[1]Department of Physics, University of California at Berkeley, Berkeley, California, USA

[2] Materials Sciences Division, Lawrence Berkeley Laboratory, Berkeley, California, USA

[*]E-mail: zhyb@berkeley.edu, crommie@berkeley.edu

**The honeycomb lattice of graphene is a unique two-dimensional (2D) system where the quantum mechanics of electrons is equivalent to that of relativistic Dirac fermions[1, 2]. Novel nanometer-scale behavior in this material, including electronic scattering[3, 4], spin-based phenomena[5], and collective excitations[6], is predicted to be sensitive to charge carrier density. In order to probe local, carrier-density dependent properties in graphene we have performed atomically-resolved scanning tunneling spectroscopy measurements on mechanically cleaved graphene flake devices equipped with tunable back-gate electrodes. We observe an unexpected gap-like feature in the graphene tunneling spectrum which remains pinned to the Fermi level ($E_F$) regardless of graphene electron density. This gap is found to arise from a suppression of electronic tunneling to graphene states near $E_F$ and a simultaneous giant enhancement of electronic tunneling at higher energies due to a phonon-mediated inelastic channel. Phonons thus act as a "floodgate" that controls the flow of tunneling electrons in graphene. This work reveals important new tunneling processes in gate-tunable graphitic layers.**

Graphene provides an ideal platform for the local study of high mobility 2D electrons because it can be fabricated on top of an insulating substrate. The availability of a back-gate electrode makes graphene the first gate-tunable 2D system directly accessible to scanning probe measurement (Fig. 1a). Previous experiments have demonstrated the power of scanning tunneling microscopy (STM) to probe the local electronic structure of graphene grown epitaxially on SiC[7-9]. That system, however, cannot be easily gated, and questions remain as to the influence of the SiC substrate on the graphene layer[6, 10]. Mechanically cleaved graphene is a useful alternative to graphene grown on SiC since it can be readily gated and is less strongly coupled to the underlying substrate.

Our graphene monolayers were prepared in a similar fashion as in Ref. [11]. Monolayers of graphene were identified using an optical microscope, and were subsequently confirmed via Raman spectroscopy (Fig. 1b)[12, 13]. Gold electrodes (30 nm) were attached to graphene using direct deposition through a stencil mask for surface cleanliness. The gold contacts remained ohmic up to a source-drain voltage of 500 mV (contact resistance < 300 Ω). Heavily doped Si under the $SiO_2$ layer was used as a back gate, allowing us to vary the carrier density in the graphene (Fig. 1a). All graphene samples were annealed at 400 ºC in ultra high vacuum (UHV) for ~ 10 hours to ensure surface cleanliness before STM measurements.

Our STM measurements were carried out in an Omicron LT-STM at low temperature (T = 4.8 K) and in a UHV environment with base pressure < $10^{-11}$ mbar. STM measurements were conducted with chemically etched metal STM tips made of tungsten or platinum/iridium alloy. To ensure that our STM tips were free of anomalies in their electronic structure, we calibrated the tips by performing tunneling differential conductance (dI/dV) measurements on a clean Au(111) surface both before and after graphene measurement. dI/dV spectra were measured through lock-in detection[9].

The STM topography of a graphene flake device is shown in Fig. 2a. Corrugations having lateral dimension of a few nanometers and vertical dimension of ~ 1.5 Å (rms value over a 60 × 60 $nm^2$ area) are observed, likely due to roughness in the underlying $SiO_2$ surface and/or intrinsic ripples of the graphene sheet[14-18]. The graphene honeycomb lattice can be clearly resolved on top of the surface corrugation, as seen more clearly in Fig. 2b.

We explored the local electronic structure of these graphene flake devices via dI/dV measurements at zero gate voltage, as shown in Fig. 2c. Strikingly, the spectrum shows a ~ 130 mV gap-like feature centered at the Fermi energy, $E_F$, as opposed to the linear density of states that might be expected from a linear band structure. This is reminiscent of a "soft gap" feature observed on SiC-grown graphene[9]. A local minimum in the tunneling conductance spectrum can also be seen at $V_D$ = -138 mV, making the spectrum asymmetric about $E_F$. Close examination of the low bias spectrum (Fig. 2c inset) reveals that the tunneling conductance does not go to absolute zero in the gap region. These observations were reproduced on more than 8 different flake devices, and the gap feature was observed in every spectrum acquired with calibrated STM tips (more than 30 different tips). The gap feature was seen to be independent of location on a sample, but the energy position of the adjacent dip feature, $V_D$, varied with tip location.

Changing the device gate voltage, $V_g$, causes the graphene Dirac point ($E_D$) to shift energetically relative to $E_F$, inducing a 2D charge carrier density of $n = \alpha V_g$ where $\alpha = 7.1 \times 10^{10}\ cm^{-2}V^{-1}$ (estimated using a simple capacitor model[1, 2]). STM dI/dV spectra taken at the same location on graphene with gate voltages ranging from $V_g$ = -60 V to $V_g$ = +60 V are shown in Fig. 3a. The width and energy position of the central gap feature do not show any dependence on gate voltage, but the conductance minimum at $V_D$ shifts monotonically with gate voltage, and even switches polarity (red arrows, Fig. 3a).

In order to further characterize the mysterious central gap feature, we measured both the temperature dependence and the effective barrier for electrons tunneling into graphene. As shown in Fig. 4a, no significant temperature dependence is observed in graphene dI/dV spectra measured at $T$ = 4.8 K and $T$ = 77.5 K (this lack of temperature dependence was observed for -60 V < $V_g$ < +60 V). Tunnel barrier measurements were carried out by measuring the STM tunnel current as a function of tip-sample separation ($z$) at constant bias for voltages inside and outside of the gap, and identical measurements were also performed on a gold sample using the same tip as a calibration. STM tunnel current depends exponentially on $z$, $I \propto e^{-z/\lambda}$, and the inverse of the decay length ($\lambda^{-1}$) gives a measure of the effective tunnel barrier. $\lambda^{-1}$ was obtained by fitting $I$ versus $z$ measurements with exponentials[19]. As seen in Fig. 4b, $\lambda^{-1}$ for graphene is comparable to that observed in the gold calibration data for biases outside of the gap (~ 2 Å$^{-1}$), but dramatically rises to a value nearly twice as large for biases within the gap (~ 4 Å$^{-1}$).

How do we interpret the anomalous graphene energy gap behavior and gate-voltage-dependent conductance minima? A consistent picture emerges if we consider phonon-mediated inelastic tunneling of electrons into the graphene flake accompanied by a strong suppression of elastic tunneling at $E_F$. In what follows we first establish a general inelastic origin of the central gap feature and we then discuss how our data can be explained by a novel phonon-based inelastic excitation mechanism.

STM tunnel current is generally enhanced if the bias voltage is high enough to provide tunneling electrons with enough energy to induce excitations that have some threshold energy $\hbar\omega_0$ (see Fig. 4c)[20]. This opens of a new inelastic tunneling channel at bias voltages of $\pm\hbar\omega_0 / e$, causing steps in dI/dV spectra that are symmetric around $E_F$ and lead to a gap-like feature with width $2\hbar\omega_0$. In the case of our graphene measurements $\hbar\omega_0$ = 63 ± 2 meV. Strong evidence for such an inelastic tunneling mechanism in graphene (as opposed to some other gap-inducing mechanism) can be seen in our data by analyzing the gate-voltage dependence of the conductance minimum observed at $V_D$. If we assume that this minimum arises from inelastic tunneling to the graphene Dirac point (a minimum in the density of states), then its energy location in our data ($eV_D$) should be offset by $\hbar\omega_0$ from its true energy location, $E_D$, in the graphene band structure (since each inelastically tunneling electron loses energy $\hbar\omega_0$):

$$E_D = e|V_D| - \hbar\omega_0 \qquad (1)$$

Because $V_D$ depends on gate voltage, this inelastic relation allows us to directly plot $E_D$ versus $V_g$, as shown in Fig. 3b. Identification of the conductance minimum at $V_D$ with the Dirac point energy, $E_D$, can be confirmed by fitting this plot with the expected dependence $E_D = \hbar v_F \sqrt{\pi\alpha |V_g - V_0|}$ as prescribed by the graphene linear band structure[1, 2]. Here $v_F$ is the Fermi velocity of graphene and $V_0$ is the shift of the Dirac point (in terms of gate voltage) due to substrate doping. An excellent fit is obtained with $v_F = 1.10 \pm 0.01 \times 10^6$ ms$^{-1}$ and $V_0 = -4.1 \pm 0.2$ V, values that are consistent with previous studies[1, 2, 21-23]. This unambiguously identifies the observed conductance minimum at $V_D$ as the Dirac point and simultaneously verifies a general inelastic tunneling origin for the central gap feature.

The question next arises as to what specific type of inelastic excitation we are observing that has energy $\hbar\omega_0 \approx 63$ meV. This excitation can be attributed to the 67 meV out-of-plane acoustic graphene phonon modes located near the **K**/**K’** points in reciprocal space[24]. Electrons with energy less than this phonon threshold energy tunnel elastically into graphene at $E_F$ (near the **K** point) with only a low probability due to suppression of electronic tunneling into states with large wavevector[25]. Once the threshold bias voltage $\pm\hbar\omega_0 / e$ is reached, however, tunneling into **K** point states is dramatically enhanced (seen as more than a factor of 10 increase in tunneling conductance) by the opening of a new inelastic channel. In this new mechanism an electron first tunnels into graphene states near the **Γ** point in reciprocal space (a virtual transition as shown in Fig. 4d) before falling into an available **K** point state via the emission of a **K’** point phonon (to conserve crystal momentum and energy). This phonon-mediated inelastic tunneling process, which involves momentum-conserving virtual transitions between 2D electronic bands, is distinctly different from previously studied inelastic tunneling in single molecules[20, 26] or localized spins[27] where momentum is not a well-defined quantity due to a lack of translational symmetry (phonon-induced inelastic tunneling in single molecules typically leads to conductivity changes only on the order of 1% in contrast to the factor of 10 seen here[26]). It is also different from band structure-dependent tunneling in silicon[28] since the wavevector dependence seen here is a result of inelastic excitations that couple the unique electronic band structure and phonon spectrum of graphene.

This mechanism is strongly supported by our observed wavefunction spatial decay rates. Within the observed energy gap (i.e. at energies below the inelastic threshold) electrons have to tunnel directly into graphene states having large crystal momentum parallel to the surface (***k***$_{//}$ = **K** or **K'**). Such states tend to decay rapidly in the vacuum region above a surface, since their evanescent local density of states (LDOS) is expected to fall off as[25]

$$LDOS(z,k_{//}) \propto e^{-z/\lambda}, \quad \lambda^{-1} = 2\sqrt{2m\phi/\hbar^2 + k_{//}{}^2} \qquad (2)$$

where $\phi$ is the workfunction and m is the mass of an electron. This accounts for the strong suppression of tunneling conductance within the energy gap region and leads to the large inverse decay length observed in the low-bias elastic tunneling channel. At voltages outside of the gap (i.e. at energies above the inelastic threshold), however, electrons tunnel via virtual excitations to states near the **Γ** point which have $k_{//} \approx 0$. Such states tend to extend further into the vacuum, leading to a smaller decay length. This can also be seen as a property of symmetry-matched components (i.e., $k_{//}= 0$) of the STM tip wavefunction at the sample surface according to Eq. (2)[25]. Using the two measured decay lengths (~ 2 Å$^{-1}$ for $k_{//}= 0$ and ~ 4 Å$^{-1}$ for $k_{//}= K$) combined with Eq. (2), we are able to extract a workfunction of $\phi = 4.6 \pm 0.2$ eV and a graphene **K** point wavevector $K = 1.7 \pm 0.4$ Å$^{-1}$. These values agree well with the known work function of graphite and the known graphene **K** point wavevector $K$ = 1.7 Å$^{-1}$. The observed lack of temperature dependence in our dI/dV spectra can also be explained by the fact that the **K/K'** phonon energy is so much higher than $k_B T$ ($k_B$ is the Boltzmann constant).

It is possible to gain insight into the electron-phonon coupling strength that underlies the inelastic tunneling process observed here. The ratio of phonon-mediated tunneling conductance at energies just outside of the gap compared to elastic tunnel conductance at energies inside of the gap can be approximated as follows (see supplementary material):

$$\left(\frac{dI}{dV}\right)_{out} \Big/ \left(\frac{dI}{dV}\right)_{in} \approx \frac{1}{7.4}\left(\frac{V_{el-ph}}{E_{\sigma}}\right)^2 \exp[(\lambda_{k_{//}=0}^{-1} - \lambda_{k_{//}=K}^{-1})z] \qquad (3)$$

Here $V_{e-ph}$ is the electron-phonon coupling matrix element connecting electronic states at the **Γ** and **K** points in reciprocal space and $E_{\sigma}$ is the energy of intermediate states near the **Γ** point on the $\sigma^*$ band ($E_{\sigma}$ ~ 4 eV[29]). If we take the STM tip height as $z \approx 5$ Å (estimated from the tip-sample junction impedance ≈ 2 GΩ) and use our measured ratio of $(dI/dV)_{out}/(dI/dV)_{in} \approx 13$, then Eq. (3) allows us to extract an experimental electron-phonon coupling strength $V_{el-ph} \approx 0.4$ eV, consistent with a theoretical estimate of $V_{el-ph} \sim 0.5$ eV based on a simple tight-binding model[30]. Our experimental value of $V_{el-ph}$, however, should only be taken as a very rough estimate because the *absolute* tip-height above graphene, *z*, is not directly measured.

In conclusion, we have demonstrated the ability to measure atomically-resolved local electronic structure of a graphene flake device while changing its charge carrier density using a back gate. The resulting graphene spectra exhibit a prominent gap-like feature that arises from a unique phonon-mediated inelastic tunneling process. Phonons thus serve as a "floodgate" that promotes electronic tunneling above a threshold energy. This phenomenon accounts for previously unexplained electronic structure in graphene grown on SiC[9], and the resulting nonlinear tunneling I-V behavior will impact future graphene devices that employ electron tunneling processes.

**Acknowledgement**

We thank D.-H. Lee, S. Louie, J. Moore, C. H. Park, G. Samsonidze, D. Wegner, L. Berbil-Bautista and C. Hirjibehedin for helpful discussions. This work was supported by DOE under contract No. DE-AC03-76SF0098. Y.Z. and F.W. acknowledge postdoctoral fellowships, and A.Z. a professor fellowship, from the Miller Institute, UC Berkeley.

**Figure Captions:**

**Figure 1: Local probe geometry of gated graphene flake device. a**. Optical image of a mechanically cleaved gated graphene flake (dark triangle in center) accessed by STM tip (tip is sketched). The graphene is contacted by gold electrodes and a back-gate voltage ($V_g$) is applied to the underlying Si substrate. **b**. Raman spectrum of the graphene sample in **a** shows a single peak at wave number ~ 2700 $cm^{-1}$, a clear signature for monolayer graphene.

**Figure 2: Graphene surface topography and differential conductance (dI/dV) spectrum. a**. Constant current STM topograph (1 V, 50 pA) of a graphene flake above a $SiO_2$ substrate. **b**. Close-up constant current STM topograph (0.15 V, 40 pA) of the graphene honeycomb lattice. **c**. dI/dV spectrum of graphene at zero gate voltage. The spectrum was acquired at a nominal junction impedance of 5 GΩ (0.5V, 100 pA). The gap width and the adjacent conductance minimum location (at $V_D$) were not sensitive to STM tip height over an impedance range of 1GΩ to 100GΩ. The inset displays a high resolution dI/dV spectrum emphasizing the central gap-like feature.

**Figure 3: Gate voltage dependence of graphene tunneling spectra. a**. dI/dV spectra taken at the same point on the graphene surface for different gate voltages, $V_g$. Spectra are acquired at the same nominal junction impedance, 5 GΩ (0.5V, 100 pA). Curves are vertically displaced for clarity. Red arrows indicate the gate-dependent positions of the adjacent conductance minimum, $V_D$, outside of the gap feature. **b**. Energy position of the Dirac point, $E_D$, as a function of applied gate voltage (extracted from the conductance minimum in **a** using Eq. (1): $E_D = e|V_D| - \hbar\omega_0$). $E_D$ and its associated uncertainty (which is smaller than the size of the symbols) are obtained by fitting the minimum with polynomials. Black curve is a fit to the data using the square root dependence of $E_D$ on gate voltage. Sketches depict linear graphene energy bands when a gate voltage is applied (filled states are orange).

**Figure 4: Characterization of the Fermi level gap feature and inelastic electron tunneling mechanism.** **a**. Graphene dI/dV conductance spectra measured at $T$ = 4.8 K and $T$ = 77.5 K (no significant temperature dependence is observed). **b**. Tunnel current inverse decay length of graphene and Au(111) surfaces probed at different bias voltages with the same STM tip. **c**. Inelastic electron tunneling due to excitation having energy $\hbar\omega_0$ . **d**. Wavevector-dependent inelastic tunneling mechanism involving graphene phonon modes near the **K** point in reciprocal space.

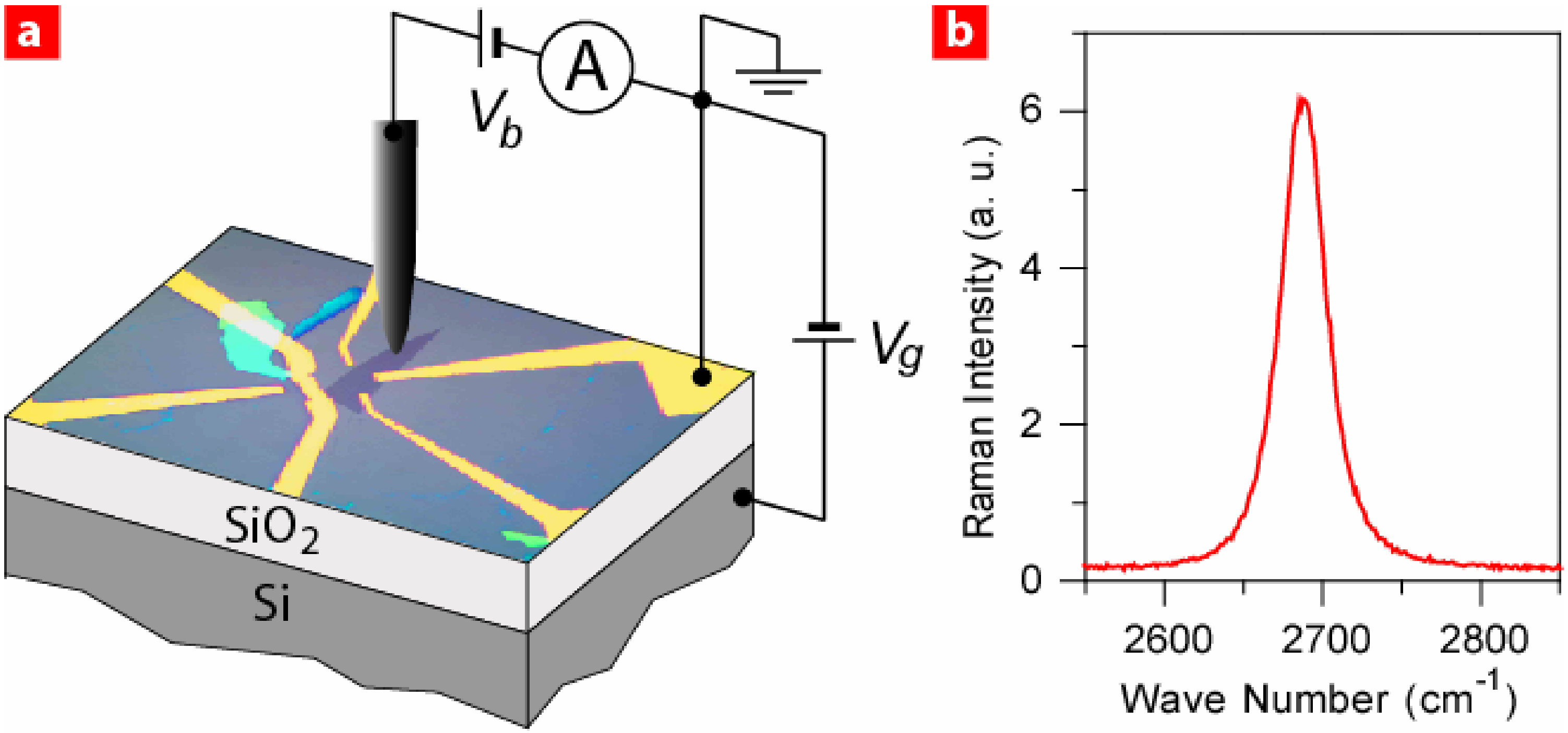

Figure 1 Y. Zhang et.al.

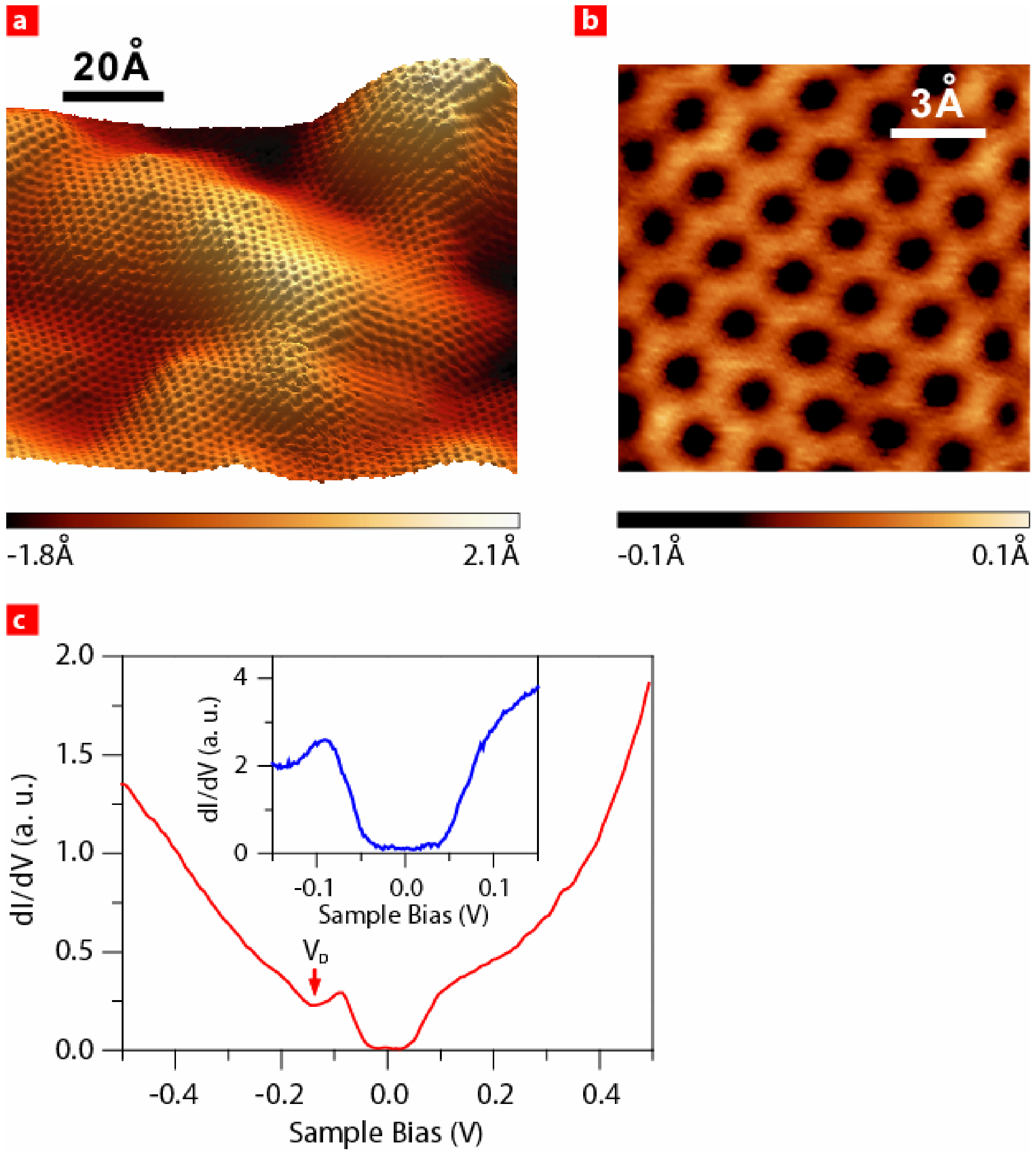

Figure 2 Y. Zhang et. al.

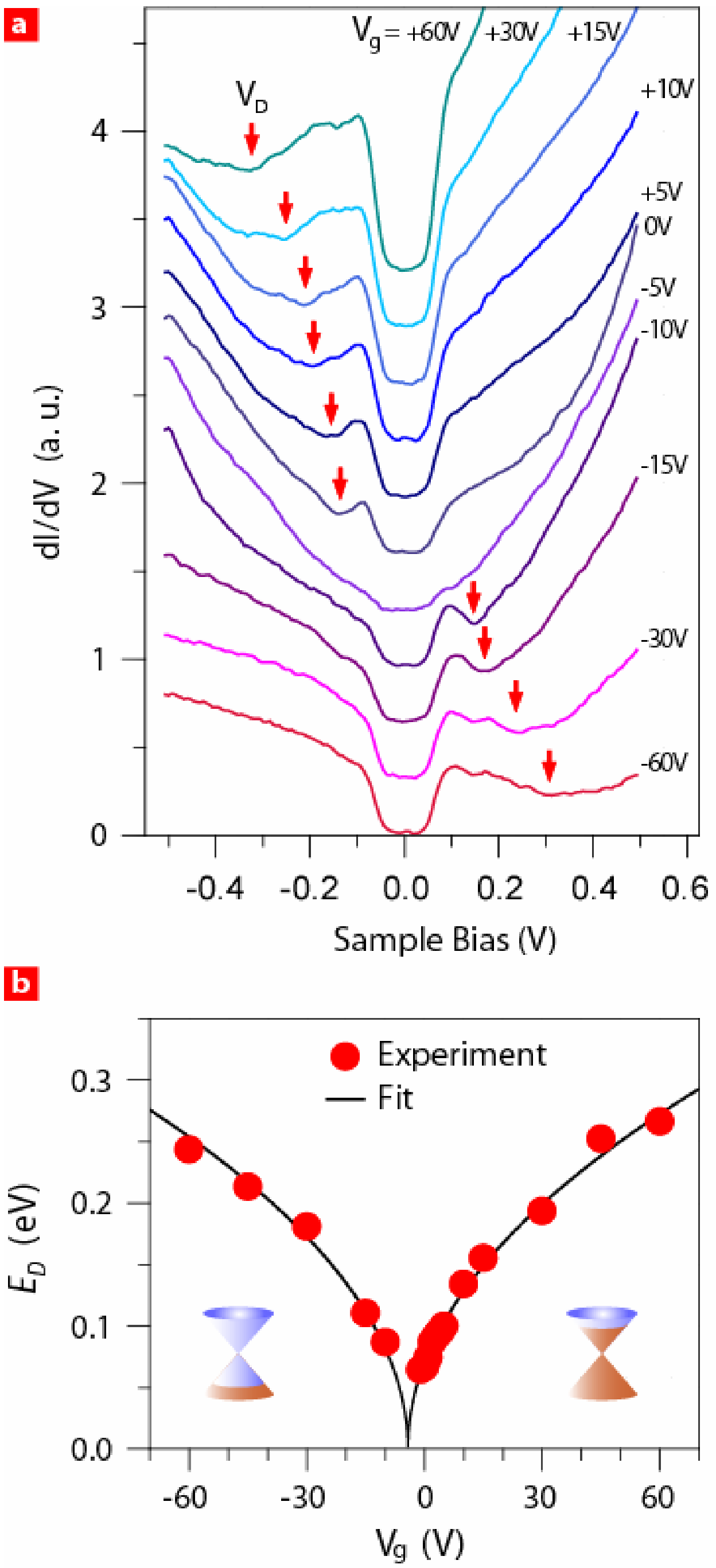

Figure 3 Y. Zhang et. al.

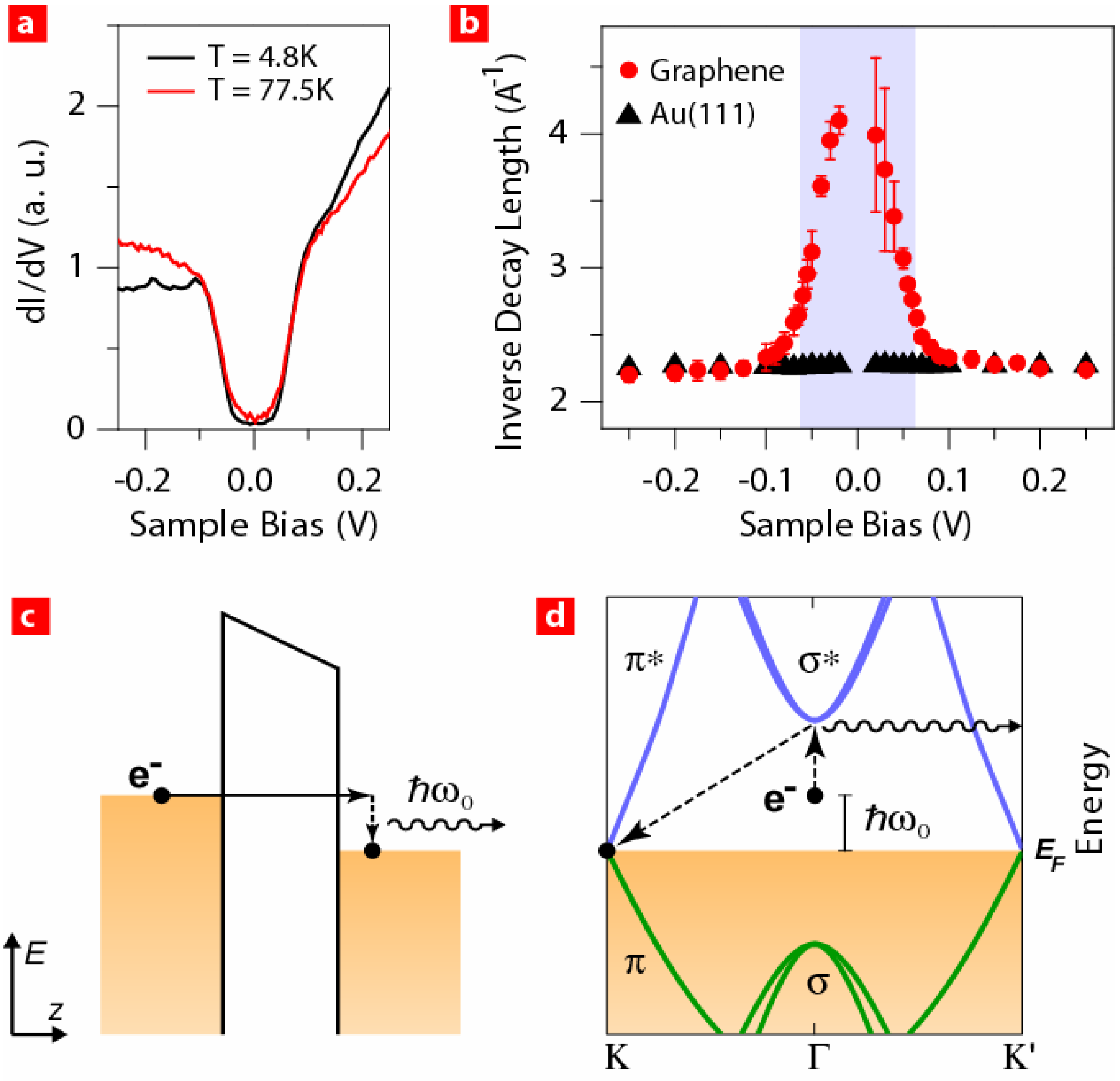

Figure 4 Y. Zhang et. al.